\documentclass[12pt,a4paper]{article}

\usepackage[english]{babel}
\usepackage[utf8x]{inputenc}
\usepackage[T1]{fontenc}

\usepackage[a4paper,top=3cm,bottom=2cm,left=3cm,right=3cm,marginparwidth=1.75cm]{geometry}

\usepackage{amsmath}
\usepackage{graphicx}
\usepackage[colorinlistoftodos]{todonotes}
\usepackage[colorlinks=true, allcolors=blue]{hyperref}
\usepackage{authblk}

\newcommand{\beq}{\begin{equation}}
\newcommand{\eeq}{\end{equation}}

\title{The dimensionless dissipation rate and the Kolmogorov (1941)
hypothesis of local stationarity in freely decaying isotropic turbulence}
\author[]{W.D. McComb}
\author[]{R.B. Fairhurst}

\affil{SUPA, School of Physics and Astronomy, University of Edinburgh,
James Clerk Maxwell Building, The King's Buildings, Edinburgh EH9 3JZ,
United Kingdom}

\begin{document}

\maketitle

\begin{abstract}
An expression for the dimensionless dissipation rate was derived from
the K\'{a}rm\'{a}n-Howarth equation by asymptotic expansion of the
second- and third-order structure functions in powers of the inverse
Reynolds number. The implications of the time-derivative term for the
assumption of local stationarity (or local equilibrium) which underpins
the derivation of the Kolmogorov `4/5' law for the third-order structure
function were studied. It was concluded that neglect of the
time-derivative cannot be justified by reason of restriction to certain
scales (the inertial range) nor to large Reynolds numbers. In principle,
therefore, the hypothesis cannot be correct, although it may be a good
approximation. It follows, at least in principle, that the quantitative
aspects of the hypothesis of local stationarity could be tested by a
comparison of the asymptotic dimensionless dissipation rate for free
decay with that for the stationary case. But in practice this is
complicated by the absence of an agreed evolution time $t_e$ for making
the measurements during the decay.  However, we can assess the
quantitative error involved in using the hypothesis by comparing the
exact asymptotic value of the dimensionless dissipation in free
decay \emph{calculated on the assumption of local stationarity} to the
experimentally determined value (e.g. by means of direct numerical
simulation), as this relationship holds for all measuring times. Should
the assumption of local stationarity lead to significant error, then
the `4/5' law needs to be corrected. Despite this, scale invariance in
wavenumber space appears to hold in the formal limit of infinite
Reynolds numbers, which implies that the `-5/3' energy spectrum does not
require correction in this limit.
\end{abstract}

\thispagestyle{empty}

\newpage
\tableofcontents
\newpage

\section{Introduction}

In this paper we extend the techniques used by McComb
\emph{et al} \cite{McComb15a}, to calculate the dimensionless
dissipation rate for stationary isotropic turbulence, to the case of
free decay. In the process we are able to shed some light on aspects of
the Kolmogorov (1941) theory which remain controversial. We begin with a
brief reconsideration of this theory.

As is well known, Kolmogorov's theory was put forward in the context of
turbulence in general. He argued that the result of the cascade would be
that one could consider the turbulence to be \emph{locally} homogeneous,
\emph{locally} isotropic; and, in time-varying situations,
\emph{locally} stationary. Regions in which this could hold, would be
restricted to a range of scales, and would necessarily be remote from
boundaries,  Since that time, it has become usual to study turbulence
which is \emph{globally} homogeneous, isotropic and indeed stationary.
This work belongs to the topic of mathematical physics, where the
problem is well-posed, and its applicability to any particular
situation, be it computer simulation or laboratory experiment, requires
some further consideration. 

In his first paper \cite{Kolmogorov41a}, Kolmogorov derived the well known
expression for the second-order structure function,
\begin{equation}
S_2(r)=C_2 \varepsilon^{2/3}r^{2/3},
\eeq
where $C_2$ is a constant. This work relied on the formulation of two
similarity principles, followed by dimensional analysis. We will refer to
it as K41A for conciseness. In his second paper \cite{Kolmogorov41b},
his starting point was the K\'{a}rm\'{a}n-Howarth equation, from which
he derived the inertial-range expression for the third-order structure
function as:
\beq
S_3 = -\frac{4}{5}\varepsilon r.
\eeq
We will refer to this as K41B. We shall return to a fuller consideration
of the `4/5' law in a later section. Here, for completeness, we note that
the `2/3' law is perhaps better known in its spectral form:
\beq
E(k)= \alpha \varepsilon^{2/3}k^{-5/3},
\eeq
where $k$  is the wavenumber and $\alpha$ is the famous Kolmogorov
constant. This result was first given by Obukhov \cite{Obukhov41}, but
that was based on a closure approximation. The first generalisation of
K41A to the spectral case appears to have been due to Onsager
\cite{Onsager45} in 1945.

After these preliminaries, the rest of the paper is organised into the
following sections:
\begin{description}
\item[Section 2] A review of recent work  on the dimensionless dissipation rate. 

\item[Section 3] We use the KHE to derive expressions for the dissipation rate in
both stationary and freely-decaying turbulence. This involves the
dimensionless KHE, as expressed in terms of dimensionless structure functions.

\item[Section 3] We revisit the derivation of the `4/5' law, with the
assumption of local stationarity, and consider the implications of this
assumption for the calculation of the asymptotic dimensionless
dissipation rate.

\item[Section 4] We use asymptotic expansions in inverse powers of the
Reynolds numbers to derive the an expression for the dimensionless
dissipation rate for free decay, and compare this to our previous result
for stationary turbulence \cite{McComb15a}.

\item[Section 5] We consider the implications for the `4/5' law and
discuss various investigations of the effect of finite viscosity and the
retention of the time-derivative term, which generally conclude that it should
be recovered in the limit of large Reynolds numbers. In contrast, our
analysis suggests that this cannot be true as the time-derivative does
not depend on either scale or Reynolds number.

\item[Section 6] We then consider the implications of our analysis for
the `2/3' and `-5/3' laws.

\end{description}

\section{The dimensionless dissipation rate}
There continues to be much interest in the fundamentals of turbulent
dissipation, as characterised by the mean dissipation rate:
\begin{equation} \label{mean-dissipation-rate}
\varepsilon(t)= \frac{\nu}{2} \sum_{\alpha,\beta=1}^{3} \left\langle
\left(\frac{\partial u_{\alpha}(t)}{\partial x_{\beta}}+\frac{\partial
u_{\beta}(t)}{\partial x_{\alpha}}\right)^2 \right\rangle,
\end{equation}
where $\nu$ is the kinematic viscosity, $u_\alpha$ is a component of the
velocity field $\mathbf{u}(\mathbf{x},t)$ expressed in cartesian tensor
notation, with the indices taking the values $\alpha,\beta = 1,\, 2,\,
\mbox{or}\, 3$. The angle brackets $\langle\dots\rangle$ denote the
operation of taking an ensemble average. For isotropic turbulence, this reduces to the
form:
\begin{equation} \label{isotropic-dissipation-rate}
\varepsilon(t)= \nu \sum_{\alpha,\beta=1}^{3} \left\langle
\left(\frac{\partial u_{\alpha}(t)}{\partial x_{\beta}}\right)^2
\right\rangle.
\end{equation}
Most of this work is based on the expression
\begin{equation}\label{TaylorSurrogate}
\varepsilon = C_\varepsilon \frac{U^3}{L},
\end{equation}
which was proposed on dimensional grounds by
Taylor in 1935 as an approximate form for the dissipation rate
\cite{Taylor35}, where $U$ is the root-mean-square velocity
and $L$ is the integral scale. Many workers in the field refer to this
as the \emph{Taylor dissipation surrogate}. However, there is a growing
tendency to rearrange it as 
\begin{equation}\label{dimensionless-dissipation}
C_\varepsilon=\frac{\varepsilon}{U^3/L},
\end{equation}
and work with the dimensionless dissipation rate $C_\varepsilon$.

As early as 1953, Batchelor \cite{Batchelor53} (in the first edition of
this book) presented evidence to suggest that $C_\varepsilon$ tends to a
constant value (nowadays denoted by $C_{\varepsilon,\infty}$), with
increasing Reynolds number. However, the growth of activity in this
topic stems from the seminal papers of Sreenivasan
\cite{Sreenivasan84,Sreenivasan98}, who established that for grid
turbulence $C_\varepsilon$ became constant for Taylor-Reynolds numbers
greater than about 50. This has inspired numerous papers reporting
experimental (including numerical) studies of the dependence of the
dissipation on Reynolds number. An account of this work, with many
references, can be found in Chapter Seven of the book by McComb
\cite{McComb14a}. 

Attempts to establish a theoretical relationship between the
dimensionless dissipation rate and the Reynolds number have been based
on the K\'{a}rm\'{a}n-Howarth equation \cite{vonKarman38} (or KHE, for short).
Lohse \cite{Lohse94} used a mean-field closure of the KHE to obtain an
approximate expression for the dependence of $C_\varepsilon$ on
$R_\lambda$, whereas Doering and Foias \cite{Doering02} established
upper and lower bounds which must be satisfied by any such relationship.

More recently McComb \emph{et al} \cite{McComb15a}, starting from the
KHE with forcing, have used an asymptotic expansion of the
structure functions in inverse powers of the Reynolds number, 
leading to a rigorous form for $C_{\varepsilon,\infty}$ and, with the aid
of numerical simulations, obtained a relationship of the form
$C_\varepsilon = C_{\varepsilon,\infty} + C/R_L + O(1/R^2_L)$, where $C$
is a constant and $R_L$ is the Reynolds number based on the integral
scale. This result is for stationary turbulence. Djenidi \emph{et al}
\cite{Djenidi17} also took the KHE as their starting point and invoked
the concept of self-preservation to assess the dependence of
dimensionless dissipation on Reynolds number. An interesting feature of
this work is their discussion at the end of Section One of its
potential importance in applications. Their analysis is for decaying
turbulence.

In this article, we extend the analysis of McComb \emph{et al}
\cite{McComb15a} to the case of freely decaying turbulence. We obtain
rigorous results in the limit of large Reynolds numbers and explore the
implication of the time-dependence for Kolmogorov's hypothesis of
\emph{local stationarity} (also known as \emph{local equilibrium}
\cite{Batchelor53}). This is a topic of continuing concern in turbulence
research: see, for instance, the recent paper by George \cite{George14}.
However, we do not present a general expression for the dependence of
the dimensionless dissipation on Reynolds number, as this requires
numerical validation. We have no \emph{a priori} way of truncating the
asymptotic expansion at small values of the Reynolds number. In the
stationary case \cite{McComb15a} that was done by comparison with the
results of a numerical simulation. For the freely decaying case that
requires further work.

\subsection{The choice of an evolution time $t_e$ in free decay}

Both stationary and freely decaying isotropic turbulence can be thought
of as initial value problems in mathematical physics. At time $t=0$ the
velocity field is chosen to be a random function of the spatial
coordinates and to have a Gaussian probability distribution. It is
characterised by an arbitrarily chosen energy spectrum that is confined
to small wavenumbers. Then, in numerical simulation of such problems,
the coupling term of the Navier-Stokes equations will lead to the energy
spreading out to higher wavenumbers, and quantities such as skewness,
energy flux and dissipation will rise to some characteristic value such
that the turbulence can be said to be evolved. The question then arises:
what is the evolution time $t_e$ for a given simulation?

For forced turbulence this is not a difficult question to answer. In
practice, the energy is found to fluctuate about a mean value, with
fluctuations in the dissipation rate lagging behind: e.g. see figure 3
in McComb \emph{et al} \cite{McComb01a}. The mean value of the energy is
determined by the energy input rate from the stirring forces and at this
stage the turbulence can be regarded as stationary. Choosing a value for
$t_e$ is simply a matter of carrying on the simulation until one
achieves satisfactory statistics.

In the case of free decay, we face the obvious problem that we cannot
simply carry on the simulation, because the turbulence is dying away.
For the most part, free decay has been studied in the context of an
assumed power-law dependence of the total energy on time. For this
reason, the onset of power-law behaviour is often taken as the criterion
for the turbulence to be well developed: see, for example, Chapter 7 of
the book \cite{McComb14a}. Although, the onset of power-law decay is a
traditional criterion for the flow to be well-developed and this was
used by, for exemple, Wang \emph{et al} \cite{Wang96} and Bos \emph{et
al} \cite{Bos07}, it does occur rather late in the decay and so there is
a possibility of choosing other criteria. For example, Fukayama \emph{et
al} \cite{Fukayama00} used the peak value of the dissipation rate as a
criterion for choosing $t_e$. But unfortunately the dissipation does not
have a peak with time when results are taken for low Reynolds numbers.
This regions is of crucial importance in establishing the curve of
dissipation against Reynolds number.

A study of the effects on the asymptotic dissipation rate of adopting
different criteria for the evolution time $t_e$ has been made by Yoffe
\cite{Yoffe12}. (This thesis can be downloaded from arXiv:1306.3408v1
[physics.flu-dyn].) As well as considering the traditional method of
taking the onset of power-law behaviour, Yoffe studied the effects of
the following criteria:
\begin{description}
\item[$t_{s}$] The time taken for the skewness to reach its peak value.
\item[$t_{\Pi}$] The time taken for the inertial transfer rate to reach
its peak value.
\item[$t_{\varepsilon}$] The time taken for the dissipation rate to reach
its peak value.
\item[$t_{\varepsilon|\Pi}$] A composite time equal to
$t_{\varepsilon}$, if peak $\varepsilon$ exists; but equal to $t_\Pi$
otherwise. 
\end{description}

We may briefly summarise these results as follows. Note that the curve
of $C_{\varepsilon}$ versus the Taylor-Reynolds number $R_{\lambda}$ was
used as a standard of comparison in assessing these results. Both $t_S$
and $t_\Pi$ took values of less than one eddy turnover time and, when
used as criteria, led to dissipation curves which fell off more rapidly
than usual, and implied an asymptotic value of
$C_{\varepsilon,\infty}(t_e)=0$. For sake of completeness, Yoffe also
took four values of evolution time in the range $3.0 \leq t_e \leq 30$,
which corresponded to the power-law regime, with time measured in units
of initial eddy turnover time. The resulting values of $C_{\varepsilon}$
clustered together quite well and lay about 50\% above the results for
forced turbulence at lower Reynolds numbers. At higher Reynolds numbers
(i.e. $R_{\lambda} \geq 50$ the results suggested asymptotic behaviour
converging on the curve for forced turbulence. 

The most interesting results were obtained from a consideration of the
dissipation rate as providing a criterion. Yoffe found that for
Taylor-Reynolds numbers below about $R_{\lambda} = 1$5, the variation of
dissipation with time did not pass through a peak, but instead seemed to
have a point of inflection. He also noticed that at these low Reynolds
numbers the inertial transfer rate appeared to go through a peak at the
same time as the dissipation passed through an inflection. On this
basis, he proposed and tested the use of a composite time
$t_{\varepsilon|\Pi}$, as listed above. He found that taking the
dissipation rate at $t=t_{\Pi}$, for $R_{\lambda}$ less than about 15, and
at $t=t_{\varepsilon}$ for larger values of the Reynolds number, led to a
continuous variation of dissipation coefficient with increasing Reynolds
number. This criterion was used in our earlier publication
\cite{McComb10b} and more recently has be found to lead to a dissipation
curve which is in good agreement with the result for forced turbulence
\cite{McComb15a} up to Taylor-Reynolds numbers of $R_{\lambda} = 358.6$
\cite{Yoffe12}.

\section{The K\'{a}rm\'{a}n-Howarth Equation (KHE)}

Our analysis is based on the KHE which, in terms of the second-
and third-order structure functions, may be written as \cite{McComb14a}: 
\begin{equation} \label{KHE-decay-C_epsilon-1}
-\frac{3}{2}\frac{\partial U^2}{\partial t}
+\frac{3}{4}\frac{\partial S_2(r,t)}{\partial t} = - \frac{1}{4r^4}
\frac{\partial}{\partial r} 
	\left(r^4S_3(r,t)\right)
+\frac{3 \nu}{2r^4} \frac{ \partial}{\partial r} 
	\left(r^4 \frac{\partial S_2(r,t)} {\partial r}\right) -I(r),
\end{equation}
where the structure functions are defined by
\begin{equation}\label{Sn}
S_n(r,t)=\left\langle[u(r,t)-u(0,t)]^n\right\rangle,
\end{equation}
and $I(r)$ represents an input term. In the main part of the present
work, we will concentrate on free decay, so the input term must then be
set equal to zero. However, at this point we will briefly consider both
free decay and forced stationary turbulence, in order to facilitate
later comparisons. We will also find it convenient to introduce new
symbols (e.g see \cite{McComb14a,McComb15a}), $\varepsilon_D(t)$ for the
energy decay rate, thus:
\begin{equation}
\varepsilon_D(t) =
-\frac{3}{2}\frac{\partial U^2(t)}{\partial t};
\end{equation}
and $\varepsilon_W$ for the energy injection rate from forcing. A
discussion of the relationship of the latter quantity to the stirring
forces used in an accompanying numerical simulation may be found in
\cite{McComb15a}, but as we do not report any numerical simulations here
we will not pursue that aspect. 

It should be noted that the KHE does not
actually contain the dissipation rate as such. We may introduce it by
means of the identity
\begin{equation}
-\varepsilon_D = -\varepsilon + \varepsilon_W
\label{identity}
\end{equation}
as derived from the Lin equation, which is the equivalent of the
KHE in wavenumber space \cite{McComb15a}. Evidently, for the stationary
case, this identity becomes $\varepsilon = \varepsilon_W$; while, for free decay,
it becomes $\varepsilon_D = \varepsilon$.

For completeness, we state the stationary case of the KHE as:
\begin{equation} \label{KHE-stat}
\varepsilon = \varepsilon_W = - \frac{1}{4r^4}
\frac{\partial}{\partial r} 
	\left(r^4S_3(r)\right)
+\frac{3 \nu}{2r^4} \frac{ \partial}{\partial r} 
	\left(r^4 \frac{\partial S_2(r)} {\partial r}\right),
\end{equation}
for scales below the forcing scale, where $I(r)\to \varepsilon_W$, and
we have invoked equation (\ref{identity}). Note that for stationarity the
dissipation is equal to the injection rate, and all time derivatives
vanish. For further discussion, see reference \cite{McComb15a}. 

We shall return to this result as required, but now we concentrate on
the main work of this paper. For free decay, we set the input term and
the injection rate equal to zero and use the equivalence of the
dissipation rate and the decay rate to write 
\begin{equation} \label{KHE}
\varepsilon(t)= \varepsilon_D(t) =
-\frac{3}{4}\frac{\partial S_2(r,t)}{\partial t} 
- \frac{1}{4r^4} \frac{\partial}{\partial r} \left( r^4S_3(r,t)\right)
+\frac{3 \nu}{2r^4} \frac{ \partial}{\partial r} \left(r^4 \frac{\partial S_2(r,t)}{\partial r}\right) .
\end{equation}
Our next step is to put this equation into dimensionless form.

\subsection{The dimensionless Karman-Howarth equation for free decay}

In order to obtain the dimensionless dissipation rate we scale the
dissipation on  $U$ and $L$.  As both of these quantities depend on the
time of decay, we perform this scaling with respect to their value at
some fiducial time $t=t_e$, as discussed in Section 2.1.  We should also
note that we are considering a well-posed  mathematical initial-value
problem, which can be realised in practice to a good approximation by
direct numerical simulation. We are not discussing grid turbulence,
which is normally a stationary flow which decays in the streamwise
direction. As is well known, the description of such a flow as decaying
in time relies on a Galilean transformation \cite{McComb90a}.

We may introduce the dimensionless
structure functions $h_n(x,\tau)$ by means of the relationship:
\begin{equation}\label{dimensionless-SF}
S_n(r,t) = U^n(t_e)h_n(x,\tau),
\end{equation}
where dimensionless time, dimensionless distance and characteristic
time, respectively,
are given by:
\begin{equation} \label{dimensionless-quantities-decay}
\begin{aligned}
\tau = \frac{t}{T} \ ; 	\ \ x=\frac{r}{L(t_e)} \ ;	 \ \ T(t_e)=\frac{L(t_e)}{U(t_e)}.
\end{aligned} \end{equation}
In terms of the dimensionless structure functions, equation (\ref{KHE}) becomes 
\begin{equation}
\label{KHEh}
\varepsilon(\tau)=-\frac{3}{4}\frac{U^2}{T} \frac{\partial h_2(x,\tau)}{\partial \tau} +\frac{U^3}{L} 
\left(
- \frac{1}{4x^4} \frac{\partial x^4h_3(x,\tau)}{\partial x} 
+\frac{3}{2x^4} \frac{\nu}{LU} \frac{ \partial}{\partial x} \left(x^4 \frac{\partial h_2(x,\tau)}{\partial x}\right)
\right) 
\end{equation}
where we leave the dependence of $U$ and $L$ on $t_e$ implicit, in the
interests of simplicity. Then, with some re-arrangement, and
substituting $R_L$ for $LU/\nu$, and from (\ref{TaylorSurrogate}) for
$\varepsilon$ in terms of $C_\varepsilon$ on the right hand side of
equation (\ref{KHEh}), we obtain
\begin{equation} \label{decay-eqn}
C_\varepsilon^{decay} (\tau) = \frac{L \varepsilon(\tau)}{U^3} = \frac{3}{4}\frac{\partial h_2(x,\tau)}{\partial \tau}
- \frac{1}{4x^4} \frac{\partial} {\partial x} \bigg(x^4h_3(x,\tau)\bigg)
+\frac{3}{2x^4R_L} \frac{ \partial}{\partial x} \left( x^4 \frac{\partial h_2(x,\tau)}{\partial x} \right).
\end{equation}
Note the introduction of the superscript \emph{decay} to distinguish
this result from the stationary case which will be denoted in the present
work by the superscript \emph{stat}.

\section{Kolmogorov's `4/5' law for freely decaying turbulence}

Before we proceed to the asymptotic expansion, it is of interest to
revisit the Kolmogorov theory for the third-order structure function
$S_3$. This is the well known \emph{Kolmogorov's `4/5' law}. It can be
obtained by integrating equation (\ref{KHE}) with respect to $r$, and
taking the infinite Reynolds number limit. Then, with some rearrangement
of terms, one obtains:
\begin{equation}\label{4/5full}
\lim_{\nu \rightarrow 0}S_3(r,t) =  -\frac{4}{5} \varepsilon_D(t)r
-\lim_{\nu \rightarrow 0}\frac{4}{r^4} \int_0^r
\frac{3r'^4}{4}\frac{\partial S_2(r',t)}{\partial t}{\rm d}r'.
\end{equation}

Two points should be noted about this. First, we have written it in
terms of the decay rate $\varepsilon_D$, in order to keep in mind its
origins. But of course this is equal to the dissipation rate, which 
is how Kolmogorov expressed it. Secondly, the term involving the
time-derivative on the right hand side was dropped by Kolmogorov, who
argued that in the inertial range of scales the turbulence could be
\emph{locally stationary}. It is worth pointing out that this was not a
specific assumption made in the second paper K41B. In fact it was introduced
as part of the definition of local isotropy in K41A. For a discussion,
see Section 4.6 of the book \cite{McComb14a}. This property was later
referred to by Batchelor \cite{Batchelor53} as \emph{local equilibrium}.
With these two steps, equation (\ref{4/5full}) reduces to the familiar
form:
\begin{equation}\label{4/5}
\lim_{\nu \rightarrow 0}S_3(r) =  -\frac{4}{5} \varepsilon r,
\end{equation}
where $r$ is in the inertial range of scales. 

We now  take the limit of infinite Reynolds numbers in equation
(\ref{decay-eqn}), and \emph{continue, for the present, to make
Kolmogorov's assumption of local stationarity}. It should perhaps be
emphasised that in this discussion the concept of local stationarity
explicitly implies the neglect of the time-derivative in equation
(\ref{decay-eqn}). In this way, we obtain the asymptotic result:
\begin{equation} 
\label{C_epsilon,infinity-decay,localstat}
\lim_{\nu \rightarrow 0} {C_\varepsilon^{ decay}}
\equiv C_{\varepsilon,\infty}^{decay} 
= - \frac{1}{4x^4} \frac{\partial}{\partial x} \left( x^4
\lim_{\nu \rightarrow 0}h_3(x)\right),
\end{equation}
where we have introduced the usual notation for the asymptotic
dimensionless dissipation rate $C_{\varepsilon,\infty}$, here decorated
by the superscript \emph{decay} . We should note the important fact that,
since the left hand side of this equation does not depend on scale, the
right hand side must be constant with respect to $x$. This is, of
course, consistent with the `4/5' law, as may be readily seen by
substituting from (\ref{dimensionless-SF}) and
(\ref{dimensionless-quantities-decay}) for $h_3(x)$ in terms of $S_3(r)$
and performing the differentiation.

We may also note from \cite{McComb15a} that the equivalent result for
the stationary  case is:
\begin{equation}
\label{C_epsilon,infinity-decay,stat}
C_{\varepsilon,\infty}^{stat}= - \frac{1}{4x^4} \frac{\partial}{\partial x} \left( x^4
\lim_{\nu \rightarrow 0}h_3(x)\right),
\end{equation}
which is just equation (35) of that paper \cite{McComb15a}, with the
addition of the superscript \emph{stat}, and the notational change of
the name for the independent variable from $\rho$ to $x$. Thus we have
the result that the expression for the asymptotic dimensionless
dissipation rate for free decay is exactly the same as that for forced
stationary turbulence (albeit evaluated at some specific time)
\emph{provided that Kolmogorov's assumption of local stationarity is
correct}. Strictly Kolmogorov's theory requires one to take the limit of
large Reynolds numbers, and when we take this into account more formally
in the next section, we find that this result leads on to an unambiguous
test of the validity of the assumption of local stationarity. 

However, for later convenience, we introduce a generalisation of
equation (\ref{C_epsilon,infinity-decay,stat}), as follows. We note that,
from a purely mathematical point of view, the left hand side is just a 
name for the expression on the right hand side. So, if we choose to
generalise the structure functions to the time-dependent case, then we
may change the `name' on the left hand side accordingly, thus:
\begin{equation}
\label{Ceps-stat-t-def}
C_{\varepsilon,\infty}^{stat}(t)= - \frac{1}{4x^4} \frac{\partial}{\partial x} \left( x^4
\lim_{\nu \rightarrow 0}h_3(x,\tau)\right)= - \frac{1}{4r^4}\frac{L}{U^3} \frac{\partial}{\partial r}
\left( r^4\lim_{\nu \rightarrow 0}S_3(r,t)\right) ,
\end{equation}
where in the second equality we have restored the structure function to
its usual form.

\section{The asymptotic expansions}

The asymptotic expansion of the dimensionless structure functions in
powers of the inverse Reynolds number is discussed in reference
\cite{McComb15a}. Here, the structure functions also depend on time, but the
procedure is the same and we may write for the $n$-order (reduced)
structure function
\begin{equation}\label{exp-free-1}
h_n(x,\tau)=h_n^{(0)}(x,\tau)+\frac{1}{R_L}h_n^{(1)}(x,\tau)+\mathcal{O} \left(\frac{1}{R_L}\right)^2.
\end{equation}

Substituting from this for the second- and third-order reduced structure
functions into equation (\ref{decay-eqn}) we obtain:
\begin{equation} \label{decay-as-eqn} \begin{split}
C_\varepsilon^{decay} (\tau) =&  -\frac{1}{4x^4} \frac{\partial}{\partial x} \left( x^4h_3^{(0)}(x,\tau) \right)
- \frac{3}{4}\frac{\partial h_2^{(0)}(x,\tau)}{\partial \tau}
+\frac{1}{R_L} \Bigg\{-\frac{1}{4x^4} \frac{\partial}{\partial x} 
	\left(x^4h_3^{(1)}(x,\tau)\right) \\
& +\frac{3}{2x^4} \frac{ \partial}{\partial x} 
\left(x^4 \frac{\partial h_2^{(0)}(x,\tau)}{\partial x}  \right)
-\frac{3}{4}\frac{\partial h_2^{(1)}(x,\tau)}{\partial \tau}
\Bigg\} 
+\mathcal{O} \left(\frac{1}{R_L}\right)^2.
\end{split} \end{equation}

By analogy with the analysis in reference \cite{McComb15a}, we may write
this as:
\begin{equation}
\label{decay-diss-eq}
C^{decay}_{\varepsilon}(\tau) =C^{decay}_{\varepsilon,\infty}(\tau) - \frac{3}{4}\frac{\partial h_2^{(0)}(x,\tau)}{\partial
\tau} + \frac{C^{decay}(\tau)}{R_L(\tau)} + \mathcal{O} \left(\frac{1}{R_L}\right)^2,
\end{equation}
where the coefficients are given by 
\begin{equation}
\label{ceps-infinity}
C^{decay}_{\varepsilon,\infty}(\tau)=  -\frac{1}{4x^4} \frac{\partial}{\partial x} \left( x^4h_3^{(0)}(x,\tau) \right),
\end{equation}
and
\begin{equation}
\label{Cdecay}
C^{decay}(\tau)= -\frac{1}{4x^4} \frac{\partial}{\partial x} \left(x^4h_3^{(1)}(x,\tau)\right) 
+\frac{3}{2x^4} \frac{ \partial}{\partial x} 
\left(x^4 \frac{\partial h_2^{(0)}(x,\tau)}{\partial x}  \right)
-\frac{3}{4}\frac{\partial h_2^{(1)}(x,\tau)}{\partial \tau}.
\end{equation}

Comparison with the stationary case, i.e. equations (40) to (42) of
reference \cite{McComb15a}, shows that the dissipation relation
(\ref{decay-diss-eq}) differs from its stationary counterpart by the
presence of the time-derivative of the zero-order part of the normalised 
structure function $h_2(x,\tau)$, while the coefficient $C^{decay}(\tau)$
is of the same form as in the stationary case but has the additional
term in the time-derivative of the first-order part, that is, $h^{(1)}_2(x,\tau)$.

\subsection{Taking the limit of infinite Reynolds numbers}

Taking the infinite Reynolds number limit of each term in equation  (\ref{decay-as-eqn}) we
find that
\begin{equation} \label{lim-decay-as-eqn} 
\lim_{\nu \rightarrow 0}C_\varepsilon^{decay} (\tau) =  -\frac{1}{4x^4}
\frac{\partial}{\partial x} \left( x^4h_3^{(0)}(x,\tau) \right) -
\frac{3}{4}\frac{\partial h_2^{(0)}(x,\tau)}{\partial \tau}.
\end{equation}

If local stationarity is again assumed, all time dependences can be
dropped, and the constant
$C^{decay}_{\varepsilon,\infty}$ in equation
(\ref{C_epsilon,infinity-decay,localstat}) can be identified as:
\begin{equation} \label{C_epsilon,infinity-decay}
 \lim_{\nu \rightarrow 0}C_\varepsilon^{decay} = C_{\varepsilon,\infty}^{decay}=
- \frac{1}{4x^4} \frac{\partial}{\partial x}
	\left( x^4h_3^{(0)}(x) \right).
\end{equation}
That is, if Kolmogorov's assumption of local stationarity is valid, the
asymptotic dimensionless dissipation rate should take the same functional form for
freely decaying decaying turbulence as it does for the forced,
stationary case.

However, if local stationarity is not assumed, the dimensionless
dissipation rate instead becomes
\begin{equation} \label{C_inf^decay}
\lim_{\nu \rightarrow 0}C_\varepsilon^{decay}(\tau) \equiv 
C_{\varepsilon,\infty}^{decay}(\tau)= 
C_{\varepsilon,\infty}^{stat}(\tau) - \frac{3}{4}\frac{\partial
h_2^{(0)}(\tau)}{\partial \tau}.
\end{equation}

Note that we continue to use $C^{stat}_{\varepsilon,\infty}(\tau)$ to indicate that
the functional form of this term is the same as in the stationary case,
but that in the case of free decay its value can depend on time. Or, invoking equations
(\ref{dimensionless-SF}) and (\ref{dimensionless-quantities-decay}),
 it may be written as:
\begin{equation}\label{C^decay_inf-S}
 C_{\varepsilon,\infty}^{decay}(t)
 =
 C_{\varepsilon,\infty}^{stat}(t)
- \frac{3}{4} \frac{L}{U^3}\frac{\partial S_2^{(0)}(t)}{\partial t}
=
C_{\varepsilon,\infty}^{stat}(t) + \Delta(t),
\end{equation}
where 
\begin{equation}\label{def-deltat}
\Delta(t) = - \frac{3}{4} \frac{L}{U^3}\frac{\partial S_2^{(0)}(t)}{\partial t},
\end{equation}
is the error made by assuming local stationarity.  As before, we point
out that, since the other terms in equation
(\ref{C_epsilon,infinity-decay}) are independent of scale, the second
term on the right-hand side of this equation must also have no dependence
on $r$. We should emphasise that these conclusions apply in the
limit of infinite Reynolds numbers, as $S_2^{(0)}$ is, by definition,
that part of the structure function which does not depend on the
Reynolds number.

\section{Implications for the `4/5' law}

Since unsustained turbulence decays with time at all Reynolds numbers,
it follows that the time derivative term in equation
(\ref{C^decay_inf-S}) must satisfy the constraint
\begin{equation}\label{extra-term-sign}
- \frac{3}{4} \frac{L}{U^3}\frac{\partial S_2^{(0)}(t)}{\partial t} \geq 0.
\end{equation}
This can be seen by, for example, differentiating Kolmogorov's `2/3' law
for the second order structure function with respect to time and noting
that the decay rate is negative. From this, it is tempting to conclude
that equation (\ref{C^decay_inf-S}), taken jointly with equation
(\ref{extra-term-sign}), implies that the asymptotic dissipation rate
for free decay must be greater than, or equal to, the asyptotic rate for
stationary turbulence. However, we must bear in mind that, although
$C_{\varepsilon,\infty}^{stat}(t)$ is of the same functional form as for
the stationary case, it is being evaluated for the time-varying case of
free decay. So we should go back to the earlier idea of a fiducial time
$t_e$, and evaluate the terms of (\ref{C^decay_inf-S}) at this time.
Accordingly, we have:
\begin{equation}\label{C^decay_inf-S-te}
 C_{\varepsilon,\infty}^{decay}(t_e)
 =
 C_{\varepsilon,\infty}^{stat}(t_e)
 - \left.\frac{3}{4} \frac{L}{U^3}\frac{\partial S_2^{(0)}(t)}{\partial t} \right|_{t=t_e}.
\end{equation}

The experimental position is unclear, but there is a view that
experimental results sugggest  $C_{\varepsilon,\infty}^{decay}\geq
C_{\varepsilon,\infty}^{stat}$. The best evidence for this is the
investigation of Bos \emph{et al} \cite{Bos07}, who compared the freely
decaying and stationary cases using direct numerical simulation,
large-eddy simulation and a closure model (EDQNM). In turn, this has
implications for the assumption of local stationarity in deriving the
`4/5' law. If that is the case, then the time derivative term in
equation (\ref{4/5full}) should be retained. 

This is not a new concern. Previously, in the context of the `4/5' law,
there has been some recognition of the possible effect of the
time-derivative term when compared to the stationary case. Lindborg
\cite{Lindborg99} used a simple model (the $k-\varepsilon$ model) to
estimate the magnitude of the unsteady term. He found the term not to be
negligible. Both Antonia \& Buratini \cite{Antonia06} and Tchoufag
\emph{et al} \cite{Tchoufag12} studied the approach of $S_3/\varepsilon
r$ to 4/5 for both stationary and decaying turbulence and both found
that the onset of the 4/5 law was at a much lower Reynolds number in the
stationary case. Antonia and Burattini \cite{Antonia06} suggested that
Taylor-Reynolds numbers of $10^3$ and $10^6$ were needed for forced and
decaying turbulence, respectively.

Tchoufag \emph{et al} \cite{Tchoufag12} came to a similar conclusion.
Their DNS results for forced turbulence were found to agree quite well with
models proposed by Moisy \emph{et al} \cite{Moisy99}, thus:
\begin{equation}
\frac{S_3(r,t)}{\varepsilon r} =
\frac{4}{5}\left[1-\left(\frac{R_\lambda}{R_{\lambda0}}\right)^{-5/6}\right], 
\end{equation}
where $R_{\lambda0} \simeq 30$; while the results for free decay agreed
well with a model of Lundgren \cite{Lundgren03}:
\begin{equation}
\frac{S_3(r)}{\varepsilon r} = \frac{4}{5} - 8.45 R_\lambda^{-2/3}.
\end{equation}
These authors further concluded that the 4/5 law was recovered at
Taylor-Reynolds numbers exceeding 5,000 in the forced case and 50,000 in
the free-decay case.

We also note the recent work of Boschung \emph{et al} \cite{Boschung16}
who considered only free decay and who studied the interplay between
finite viscous effects and the unsteady term, using direct numerical
simulation and a model closure which allowed them to extend their
results up to Taylor-Reynolds numbers of $10^4$. Even at such high
Reynolds numbers, these authors found that the inertial range is quite
short. They also found that the viscous term acts as a sink of energy
while the unsteady term acts as a source, at all scales. They concluded 
that the inertial range is the region surrounding the point where these
two effects cancel out.

The implication of these investigations is that the Kolmogorov `4/5' law
should be recovered in the limit of infinite Reynolds numbers in the
case of free decay. Yet, in view of our present results, this cannot be
entirely true. As we have seen, the time-derivative term in equation
(\ref{4/5full}) does not depend on either scale or Reynolds number.
Accordingly, unless it is inherently zero for some other reason, this
is the limiting case. That is to say,  if the time derivative is
indeed non-zero, $S_3/\varepsilon r$ should never reach 4/5; or:
\begin{equation}
\label{limit}
|S_3(r,t)|< \frac{4}{5}\varepsilon(t) r \ \ \forall R.
\end{equation} 
In wavenumber space, this would suggest  that the peak flux (through
wavenumber) would never equal the dissipation rate, as previously
pointed out by Sagaut and Cambon \cite{Sagaut08} and McComb \emph{et al}
\cite{McComb10b}. This result could have implications for other aspects
of the Kolmogorov-Richardson phenomenology and we discuss these in the
next section.

\section{Implications for the Kolmogorov `2/3' and `-5/3' laws}

As noted in the Introduction, the main outcome of the Kolmogorov (1941)
theory is the `2/3' law for the second-order structure function,
$S_2(r,t)=C_2\varepsilon^{2/3} r^{2/3}$, and the corresponding energy
spectrum, $E(k,t)= \alpha \varepsilon^{2/3} k^{-5/3}$, where in both
cases the independent variables are restricted to their inertial range.
We will concentrate on the spectral case, while bearing in mind that the
real space case can be recovered by Fourier transformation. We begin by
Fourier transforming the KHE in order to obtain the Lin equation (e.g.
see reference \cite{McComb14a}). This
may be written as:
\begin{equation}
\label{lin}
\frac{\partial E(k,t)}{\partial t}= W(k) +T(k,t)-D(k,t),
\end{equation}
where $k$ is the wavenumber, $E(k,t)$ is the energy spectrum (obtained
from Fourier transformation of $S_2(r,t)$; $W(k)$ is an energy input
term, arising from stirring forces; $T(k,t)$ is the transfer spectrum;
and $D(k,t)= 2 \nu k^2 E(k,t)$ is the dissipation spectrum. An
expression for $T(k,t)$ can be found, for instance, as equation (3.14)
in the book \cite{McComb14a}, but that will not be needed here.

We can apply this equation either to stationary turbulence, with the
time derivative set equal to zero, and all time dependences dropped; or
to freely decaying turbulence, with $W(k)=0$. Alternatively, we can
write equation (\ref{lin}), with some rearrangement, in unified form for
both cases, as \cite{McComb14a}:
\begin{equation}
\label{un_lin}
-T(k,t)= I(k,t) - D(k,t),
\end{equation}
where $I(k,t)$ stands for either the time derivative term or the
stirring spectrum. In the latter case the time dependences should be
omitted in order to indicate stationarity.  Then, by comparison with
equation (\ref{decay-as-eqn}), we may deduce the form of (\ref{un_lin})
in the infinite Reynolds number limit as:
\begin{equation}
\label{un_lin_lim}
-\lim_{\nu \to 0}\left. T(k,t)\right|_{\varepsilon = const} =
\lim_{\nu\to 0}\left. I(k,t) \right|_{\varepsilon = const} -
\lim_{\nu\to 0}\left. D(k,t) \right|_{\varepsilon = const}.
\end{equation} 

We now need more explicit forms on the right hand side. The term
involving the time derivative is the Fourier transform of 
\[
\lim_{\nu\to 0}\left. I(r,t) \right|_{\varepsilon = const} =
I(r,t) = \lambda(t)
\]
where $\lambda(t)$ is independent of both scale $r$ and Reynolds number.
Hence, its Fourier transform with respect to wavenumber $k$ is just a
delta function at the origin in $\mathbf{k}$-space.

Evaluating the second term on the right hand side of equation (\ref{un_lin_lim})
is a little more tricky; but the analysis, having been given by Batchelor
\cite{Batchelor53} and developed by Edwards \cite{Edwards65}, is quite
well known. We take the limit of the Kolmogorov wavenumber $k_d$, thus:
\[
\lim_{\nu\to 0}\left. k_d \right|_{\varepsilon = const} = \lim_{\nu\to
0}\left.\frac{\varepsilon(t)}{\nu^3} \right|_{\varepsilon = const} \to \infty.
\]
Thus, in this limit, the dissipation rate must be concentrated at
$k=\infty$ and can be represented by $\varepsilon(t)
\delta(k-\infty)$. 
In all then, we can write (\ref{un_lin_lim}) as:
\begin{equation}
\label{sfe}
-\lim_{\nu \to 0}\left. T(k,t)\right|_{\varepsilon = const} =
\lambda(t)\delta(k) - \varepsilon(t)\delta(k-\infty) =
\varepsilon(t)\delta(k)-\varepsilon(t)\delta(k-\infty), 
\end{equation}
where the equality $\lambda(t) =\varepsilon(t)$ follows from integrating
both sides over all values of $k$, and invoking conservation of energy.

Equation (\ref{sfe}) was first given for stationary turbulence by
Edwards \cite{Edwards65}, in the course of testing a closure
approximation based on his self-consistent field theory. Here we see
that it is also the limiting case for freely decaying turbulence as
well. Introducing the energy flux $\Pi(\kappa,t)$ through mode
$\kappa$, by the relationship
\begin{equation}
\Pi(\kappa,t) = -\int_0^\kappa\,dk\,T(k,t),
\end{equation}
it follows from equation (\ref{sfe}) that 
\begin{equation}
\Pi(\kappa,t) =  \varepsilon(t), \quad \forall \kappa.
\end{equation}
Hence we have scale invariance, which is the necessary condition for an
inertial range. Here this applies for all values of the wavenumber, in
the limit of infinite Reynolds numbers\footnote{Strictly we should
exclude the exact values $\kappa = 0$ and $\kappa=\infty$.}. It was
argued by Edwards \cite{Edwards65} that the $-5/3$ spectrum would apply
for all wavenumbers under these circumstances. So it appears that the
unsteady term in free decay appears only as a source term in the
spectral energy balance and hence, whatever its effect on the `4/5' law,
does not affect the Kolmogorov energy spectrum.

\section{Conclusions}

Previously McComb \emph{et al} \cite{McComb15a} found that, for
stationary turbulence, the dimensionless dissipation rate took the form:
\begin{equation}
\label{final_stat}
C_\varepsilon = C_{\varepsilon,\infty} + C/R_L + \mathcal{O}(1/R_L^2): \quad \mbox{stationary case};
\end{equation}
where $C$ is a constant which depends on those parts of the second- and
third-order structure functions which are independent of the Reynolds
number. The asymptotic value and the coefficient were evaluated
respectively as $C_{\varepsilon,\infty} = 0.468 \pm 0.006$ and $C=18.9
\pm 1.3$ from the numerical simulation \cite{McComb15a}.

In the present article, we have derived the asymptotic form for the
freely decaying  case and it is tempting to go on and write the
analogous expression for free decay as:
\begin{equation}
\label{final_decay}
C^{decay}_\varepsilon(t_e) = C^{decay}_{\varepsilon,\infty}(t_e) -
\left. \frac{3}{4}\frac{L}{U^3}\frac{\partial S_2^{(0)}}{\partial t}\right|_{t=t_e} + C^{decay}(t_e)/R_L(t_e) +
\mathcal{O}(1/R_L^2): \quad \mbox{free decay}.
\end{equation}
However, there are two points at issue here.

First, we need to justify the truncation to the first-order term in
(\ref{final_decay}). At low values of the Reynolds number, it may be
necessary to take higher orders into account. For instance, with
magnetohydrodynamic turbulence, it is necessary to take the next order
of the expansion into account, although the effect is small
\cite{Linkmann15a}. However, there is quite good support from an earlier
investigation at low Reynolds numbers \cite{McComb10b} suggesting that
the first-order truncation in (\ref{final_decay}) is justified.

The second issue is the absence of an agreed criterion for choosing the
evolved time $t_e$. At present this is not an aspect which is discussed
much in this field and there appears to be a tacit acceptance that, for
times of decay greater than some evolved time, the asymptotic
dissipation rate in free decay will take some universal value. The
results of Yoffe \cite{Yoffe12} suggest that this may be true, provided
that $t_e$ is chosen to be something like three eddy turnover times, or
larger. But this means that one is losing much of the decay process by
working at long evolution times. So there is some attraction in  the use
of the composite time $t_{\varepsilon|\Pi}$, as discussed in Section
2.1. This is a matter which deserves further study.

Lastly, as seen in Section 5.1, $\Delta(t)$ as defined by equation
(\ref{def-deltat}), is the error made by assuming local stationarity in
the case of freely decaying turbulence. If we can show a physical basis
for adopting the composite form of evolution time, then the results of
Yoffe \cite{Yoffe12} indicate that $\Delta(t)$ can be evaluated from a
comparison between the asymptotic results for forced and freely decaying
turbulence. This will be the subject of further work. However, it may
also be pointed out that this error can also be found from suitable
limiting procedures applied to equation (\ref{final_decay}), as this
equation holds for all times. This would be a more cumbersome procedure
but could be carried out using conventional numerical simulations.

\section*{Acknowledgements}

One of us (WDM) would like to thank Katepalli Sreenivasan for a helpful
exchange of emails and in particular for pointing out the significance
of the `fragment' by Onsager.


\end{document}